\documentstyle[12pt]{article}

\begin{document} 

\begin{titlepage}

\hrule 
\leftline{}
\leftline{CHIBA UNIVERSITY Preprint
          \hfill   \hbox{\bf CHIBA-EP-114}}
\leftline{\hfill   \hbox{hep-th/9906129}}
\leftline{\hfill   \hbox{September 1999}}
\vskip 5pt
\hrule 
\vskip 1.0cm

\centerline{\large\bf 
Non-Abelian Stokes Theorem  and Quark Confinement} 
\vskip 0.5cm
\centerline{\large\bf  
in SU(3) Yang-Mills Gauge Theory$^*$
}
\centerline{\large\bf  
}
\centerline{\large\bf  
}
\centerline{\large\bf  
}

\vskip 1cm

\centerline{{\bf 
Kei-Ichi Kondo$^{1,2}{}^{\dagger}$
and Yutaro Taira$^2{}^{\ddagger}$
}}  
\vskip 1cm
\centerline{{\it
$^1$ Department of Physics, Faculty of Science, 
Chiba University,  Chiba 263-8522, Japan}}
\vskip 0.5cm
\centerline{{\it 
$^2$ Graduate School of Science and Technology,
  Chiba University, Chiba 263-8522, Japan}}

\vskip 0.5cm

\centerline{{\bf Abstract}}

We derive a new version of SU(3) non-Abelian Stokes theorem by
making use of the coherent state representation on the coset space
$SU(3)/(U(1)\times U(1))=F_2$, the flag space. 
Then we outline a derivation of the area law of the Wilson loop in
$SU(3)$ Yang-Mills theory in the maximal Abelian gauge
(The detailed exposition will be given in a forthcoming article). 
This derivation is performed by combining the non-Abelian Stokes theorem 
with the reformulation of the
Yang-Mills theory as a perturbative deformation of a topological field theory
 recently proposed by one of the authors. 
Within this framework, we show that the fundamental quark is confined even if
$G=SU(3)$ is broken by partial gauge fixing into 
$H=U(2)$ just as $G$ is broken to $H=U(1) \times U(1)$.  
An origin of the area law is related to the geometric phase of the
Wilczek-Zee holonomy for $U(2)$. 
Abelian dominance is an 
immediate byproduct of these results and magnetic monopole plays the dominant role in this derivation.


\vskip 0.2cm
\hrule  
\vskip 0.5cm
\leftline{$^\dagger$E-mail:  kondo@cuphd.nd.chiba-u.ac.jp} 
\leftline{$^\ddagger$E-mail: taira@cuphd.nd.chiba-u.ac.jp} 
\vskip 0.5cm
\noindent
$^*$  To be published in Mod. Phys. Lett. A.

\vskip 0.5cm  


\end{titlepage}

\pagenumbering{arabic}
\baselineskip 23pt

\section{Introduction}
\setcounter{equation}{0}
\par
Quantum Chromodynamics (QCD) is a quantized gauge field theory with
color $SU(3)$ as the gauge group.  
Usually, a simplified version of QCD with a gauge
group $SU(2)$ is first investigated to avoid technical complexity of
dealing with $SU(3)$ group.
This is also the case in the study of quark confinement in
QCD.  However, this simplification might lose some important features
which may appear only when we begin to consider $SU(3)$ case. This is
anticipated because it is generally believed that the quark
confinement is not entirely of the kinematic origin and that some
dynamical information on non-Abelian Yang-Mills gauge theory is
indispensable to lead to the complete resolution of the problem. 
\par
Recently, the dual superconductor picture \cite{DSC} of quark
confinement in QCD has been extensively investigated based on the
idea of Abelian projection \cite{tHooft81}, see \cite{review} for
review.   In this scenario, the original gauge group
$G=SU(N)$ is broken down to the maximal torus subgroup 
$H=U(1)^{N-1}$ (up to the discrete Weyl symmetry) by
the partial gauge fixing.  This will imply that the $(N-1)$ species
of Abelian magnetic monopoles will be responsible for quark
confinement ($N\ge 2$).  In the presence of the elementary scalar
field, e.g., in the Yang-Mills-Higgs theory, we can break the gauge
symmetry
$G$ to any subgroup $H$ by choosing appropriately the expectation
value of the Higgs scalar field. 
However, QCD does not have the elementary scalar field in the
theory, in sharp contrast with the supersymmetric version
\cite{SW94}.  If the scalar field would appear in QCD and carry the
degrees of freedom of magnetic monopoles, it should be provided
as a composite of gauge field.  This was indeed the case in $SU(2)$,
see
\cite{KondoI,KondoII}.  The dynamics  of true $SU(3)$ QCD may restrict
the  type of the composite scalar field as a carrier of magnetic
monopoles which is responsible for the quark confinement.  If so, the
subgroup $H$ could be restricted.
\par
In this Letter, we give a first report on the extension of the
previous analyses of quark confinement in $SU(2)$ Yang-Mills theory
performed in a series of papers
\cite{KondoI,KondoII,KondoIV,KondoVI} to
the $SU(3)$ case.  The technical details will be given in a subsequent
 paper \cite{KT99}.
Our results suggest that the fundamental quark in QCD is confined in
the sense of the area law of the Wilson loop even if $SU(3)$ is
restricted to 
$H=U(2)$ just as 
$U(1) \times U(1)$, without contradicting with the maximal Abelian
gauge.

\section{Coherent state on the flag space}
\subsection{General procedure}
\par
In this Letter we shall derive  a version of the non-Abelian Stokes
theorem (NAST) for $SU(3)$, generalizing the NAST for
$SU(2)$ derived in
\cite{DP89,KondoIV}.   Further generalization to $SU(N)$ is
straightforward
\cite{KT99}.  
First of all, we construct the coherent state corresponding to the
coset representatives $\xi \in G/\tilde{H}$.  For inputs, we prepare
(a) the gauge group $G$ and its (semi simple) Lie algebra
${\cal G}$ with the generators $\{ T^A  \}$
of the Lie algebra  being rewritten in terms of the Cartan basis 
$\{ H_i, E_\alpha, E_{-\alpha} \}$;
(b)  the Hilbert space $V^\Lambda$ as a carrier (the
representation space) of the unitary irreducible representation
$\Gamma^\Lambda$ of $G$;   
(c)  a reference state 
$|\Lambda \rangle$ within the Hilbert space $V^\Lambda$,
which can be normalized to unity,
$
 \langle \Lambda |\Lambda \rangle = 1 .
$
\par
We define the {\it maximal stability subgroup (isotropy subgroup)} $\tilde{H}$ as a
subgroup of
$G$ that consists of all the group elements $h$ that leave the
reference state $|\Lambda \rangle$ invariant up to a phase factor,
i.e.,
$
  h |\Lambda \rangle = |\Lambda \rangle e^{i\phi(h)}, h \in \tilde{H} .
$
The phase factor is unimportant in the following because we consider
the expectation value of any operators in the coherent state.
For every element $g\in G$, there is a unique decomposition of $g$
into a product of two group elements, 
$
 g = \xi h, \xi \in G/\tilde{H}, h \in \tilde{H} ,
$
for $g \in G$.
We can obtain a unique coset space for a given $|\Lambda \rangle$.
The action of arbitrary group element $g\in G$  on 
$|\Lambda \rangle$ is given by
$
  g |\Lambda \rangle = \xi h  |\Lambda \rangle
  = \xi  |\Lambda \rangle  e^{i\phi(h)}.
$
\par
The coherent state is constructed as
$
  |\xi, \Lambda \rangle = \xi |\Lambda \rangle .
$
This definition of the coherent state is in one-to-one
correspondence with the coset space $G/\tilde{H}$ and the coherent states
preserve all the algebraic and topological properties of the coset
space $G/\tilde{H}$.
If $\Gamma^\Lambda({\cal G})$ is Hermitian, then $H_i^\dagger=H_i$, 
and $E_{\alpha}^\dagger = E_{-\alpha}$.  Thus the coherent state is
given by \cite{FGZ90}
\begin{eqnarray}
  |\xi, \Lambda \rangle = \exp \left[ 
  \sum_{\beta '} \left( \eta_\beta E_\beta - \eta_\beta^* E_{-\beta}
\right) \right] |\Lambda \rangle,
\end{eqnarray}
where 
$|\Lambda \rangle$ is the highest-weight state such that 
$|\Lambda \rangle$  is
(i) annihilated by all the shift-up operators $E_{\alpha}$ with
$\alpha>0$, i.e., 
$
 E_{\alpha} |\Lambda \rangle = 0 (\alpha>0) ;
$
(ii) mapped into itself by all diagonal operators $H_i$, i.e., 
$
 H_{i} |\Lambda \rangle = \Lambda_i |\Lambda \rangle ;
$
(iii) annihilated by some shift-down operators $E_{\alpha}$ with
$\alpha<0$, not by other $E_{\beta}$ with $\beta<0$, i.e., 
$
 E_{\alpha} |\Lambda \rangle = 0 ({\rm some~} \alpha<0) ;
$
$
 E_{\beta} |\Lambda \rangle = |\Lambda+\beta \rangle 
 ({\rm some~} \beta<0) ;
$
and the sum $\sum_{\beta '}$ is restricted to those shift operators
$E_{\beta}$ which obey (iii).
\par
Two coherent states are non-orthogonal, 
$
 \langle \xi' , \Lambda | \xi, \Lambda \rangle \not= 0 ,
$
but normalized to unity, 
$
 \langle \xi, \Lambda | \xi, \Lambda \rangle = 1  .
$
The coherent state spans the entire space $V^\Lambda$.
By making use of the group-invariant measure $d\mu(\xi)$ of
$G$ which is appropriately normalized, we obtain
\begin{eqnarray}
  \int |\xi, \Lambda \rangle d\mu(\xi) 
  \langle \xi,  \Lambda |
  = I ,
\end{eqnarray}
which shows that the coherent states are complete, but overcomplete.
This resolution of identity is very important to obtain the path
integral formula given below.
\par
For concreteness, we focus on the $SU(3)$ case.  
Using the Dynkin index $[m,n]$ ($m,n$: integers), the
highest weight $\Lambda$ can be written as
$
  \vec \Lambda = m \vec h_1 + n \vec h_2
$
($m,n$ are non-negative integers for the highest weight)
where $h_1, h_2$ are highest weights of two fundamental
representations of $SU(3)$ corresponding to $[1,0], [0,1]$
respectively, i.e.,
$
  \vec h_1 = ({1 \over 2}, {1 \over 2\sqrt{3}}),
  \vec h_2 = ({1 \over 2}, {-1 \over 2\sqrt{3}}) .
$
Therefore, 
$
  \vec \Lambda =  ({m+n \over 2}, {m-n \over 2\sqrt{3}}) .
$
The maximal stability group $\tilde{H}$ is given by $\tilde{H}=U(2)$ if
$m=0$ or $n=0$ (case (I)), whereas $\tilde{H}$ is the maximal torus group 
$H=U(1) \times U(1)$ if
$m\not=0$ and $n\not=0$ (case (II)). 
Therefore, for the representation with the highest weight
$\Lambda$, the coset $G/\tilde{H}$ is given by
$SU(3)/U(2)=SU(3)/(SU(2)\times U(1))=CP^2$ in the case (I) and
$SU(3)/(U(1)\times U(1))=F_2$ in the case (II). Here, $CP^{N-1}$ is the
complex projective space and $F_{N-1}$ is the flag space
\cite{Perelomov87}. Therefore, the two fundamental representations
belong to the case (I), so the maximal stability group is $\tilde{H}=U(2)$,
rather than the maximal torus group $H=U(1) \times U(1)$.  
The implications of this fact to the mechanism of quark confinement
is discussed in the following.

\par
\subsection{Explicit form of the coherent state on flag space}
\par
The SU(2) case is well known.  The coherent state for
$F_1:=SU(2)/U(1)$ is obtained as
\begin{eqnarray}
 |j, w \rangle = \xi(w) |j, -j \rangle 
 = e^{\zeta J_{+} - \bar \zeta J_{-}}  |j, -j \rangle 
 = {1 \over (1+|w|^2)^j}e^{ w J_{+}}  |j, -j \rangle ,
\end{eqnarray}
where
$
  J_{+} = J_1+iJ_2 ,
  J_{-}=J_{+}^\dagger ,  
$
$|j, -j \rangle $ is the lowest
state, 
$|j, m=-j \rangle $ of $|j, m \rangle  ,
$
and 
$
 w = {\zeta \sin |\zeta| \over |\zeta| \cos |\zeta|}  .
$
Note that $(1+|w|^2)^{-j}$ is a normalization factor to ensure 
$
 \langle j, w |j, w \rangle = 1 ,
$
which is obtained from the Baker-Campbell-Hausdorff (BCH) formulas.
The invariant measure is given by
\begin{eqnarray}
 d\mu = {2j+1 \over 4\pi} {dwd\bar w \over (1+|w|^2)^2} .
\end{eqnarray}
For
$
J_A = {1 \over 2}\sigma^A (A=1,2,3)  
$
with Pauli matrices $\sigma^A$, we obtain
$
  J_{+} = \pmatrix{0 & 1 \cr 0 & 0} ,
$ 
and
$
  e^{ w J_{+}}  = \pmatrix{1 & w \cr 0 & 1} \in F_1 = CP^1=
SU(2)/U(1) \cong S^2.
$
The complex variable $w$ is a CP$^1$ variable written as 
$
 w = e^{-i\phi} \tan {\theta \over 2},
$
in terms of the polar coordinate on $S^2$ or Euler angles, see
\cite{KondoII}. 
\par
Now we proceed to the $SU(3)$ case.  The coherent state for 
$F_2 = SU(3)/U(1)^2$ is given by
$
 | \xi, \Lambda \rangle = \xi(w) | \Lambda \rangle 
 := V^\dagger(w) | \Lambda \rangle ,
$
with the highest(lowest)-weight state $| \Lambda \rangle$ and
\begin{eqnarray}
 | \xi, \Lambda \rangle
 = \exp \left[\sum_{\alpha=1}^{3}(\zeta_\alpha E_\alpha -
\bar\zeta_\alpha E_\alpha^\dagger ) \right] | \Lambda \rangle
=  e^{-{1 \over 2}K(w,\bar w)}  
\exp \left[\sum_{\alpha=1}^{3} \tau_\alpha E_\alpha \right]  |
\Lambda \rangle ,
\label{cohe}
\end{eqnarray}
where
$e^{-{1 \over 2}K}$ is the normalization factor obtained from
the K\"ahler potential (explained later)
\begin{eqnarray}
  K(w,\bar w) 
  &:=& \ln [(\Delta_1(w,\bar w))^m(\Delta_2(w,\bar w))^n] ,
\\
 \Delta_1(w,\bar w) &:=& 1+|w_1|^2+|w_2|^2, \quad
 \Delta_2(w,\bar w) := 1+|w_3|^2+|w_2-w_1 w_3|^2 .
\end{eqnarray}
The coherent state 
$
 | \xi, \Lambda \rangle 
$
is normalized, so that
$
 \langle \xi, \Lambda | \xi, \Lambda \rangle = 1 .
$
We can obtain the inner product \cite{KT99}
\begin{eqnarray}
 \langle \xi', \Lambda | \xi, \Lambda \rangle 
 =  e^{K(w, \bar w')} e^{-{1\over2}[K(w',\bar w')+K(w,\bar w)]} ,
 \label{norm}
\end{eqnarray}
where 
$
  K(w, \bar w') := 
 \ln [1+\bar w_1{}' w_1+\bar w_2{}'w_2]^m[1+\bar w_3{}' w_3
 +(\bar w_2{}'-\bar w_1{}'\bar w_3{}')(w_2-w_1w_3)]^n .
$
The $SU(3)$ invariant measure is given (up to a constant factor) by
\begin{eqnarray}
 d\mu(\xi) = d\mu(w, \bar w)
 = D(m,n)[(\Delta_1)^m (\Delta_2)^n]^{-2} \prod_{\alpha=1}^{3}
dw_\alpha d\bar w_\alpha , 
\end{eqnarray}
where $D(m,n)={1 \over 2}(m+1)(n+1)(m+n+2)$ is the dimension of the
representation. For
$
  E_1 := (\lambda_1-i\lambda_2)/(2\sqrt{2}), 
  E_2 := (\lambda_4-i\lambda_5)/(2\sqrt{2}), 
  E_3 := (\lambda_6-i\lambda_7)/(2\sqrt{2}), 
$
with the Gell-Mann matrices $\lambda_A(A=1,\cdots,8)$,
we obtain
$
\exp \left[\sum_{\alpha=1}^{3} \sqrt{2}\tau_\alpha E_\alpha \right]  
 = \pmatrix{1 & w_1 & w_2 \cr 0 & 1 & w_3 \cr 0 & 0 & 1}^t 
 \in F_2 = SU(3)/U(1)^2 .
$
Two sets of three complex variables are related as
$
 w_1 = \tau_1, w_2 = \tau_2 + \tau_1\tau_3/2, w_3 = \tau_3 ,
$
or conversely
$
 \tau_1 = w_1, \tau_2 = w_2 - w_1 w_3/2, \tau_3 = w_3 .
$
\par
Any element of $F_{N-1}=SU(N)/U(1)^{N-1}$ is written as an upper
triangular matrix with
$N(N-1)/2$ complex numbers.  It is not difficult to extend the above
results to $SU(N)$, see \cite{KT99}.

\section{Non-Abelian Stokes theorem for SU(3)}
\par
For the infinitesimal deviation $\xi'=\xi+d\xi$ (which is
sufficient to derive the path integral formula), we find from
(\ref{cohe}) and (\ref{norm}) 
\begin{eqnarray}
 \langle \xi', \Lambda | \xi, \Lambda \rangle
 &=& \exp ( i \omega + O((dw)^2) ) ,
 \\ \omega(x) 
 &:=& \langle \Lambda | i V(x) d V^\dagger(x) |\Lambda \rangle  
  = \langle \Lambda | i \xi^\dagger (x) d\xi(x) |\Lambda
\rangle ,
\end{eqnarray}
where $d:=dx^\mu \partial_\mu$ denotes an exterior derivative
and the one-form $\omega$ is given by
\begin{eqnarray}
  \omega = im {w_1d\bar w_1+w_2d\bar w_2 \over \Delta_1(w, \bar w)} 
  + in {w_3d\bar w_3+(w_2-w_1 w_3)(d\bar w_2-\bar w_1d\bar w_3-\bar
w_3d\bar w_1)
\over \Delta_2(w, \bar w)} ,
\end{eqnarray}
up to the total derivative.
Here the $x$-dependence of $V$ comes through that of $w(x)$ (the
local field variable $w(x)$), i.e., $V(x)=V(w(x))$.
\par
The Wilson loop operator $W^C[{\cal A}]$ is defined as the
path-ordered exponent along the closed loop $C$,
$
  W^C [{\cal A}] :=  {\rm tr} \left[ {\cal P} 
 \exp \left( ig \oint_C {\cal A}
 \right) \right]/{\rm tr}(1) ,
$
where ${\cal A}$ is the Lie-algebra valued connection one-form, 
$
 {\cal A}(x) = {\cal A}_\mu^A(x)T^A dx^\mu = {\cal A}^A(x) T^A .
$
Repeating the same
steps as those given in
\cite{KondoIV}, we obtain the path integral representation of the
Wilson loop,
\begin{eqnarray}
 W^C[{\cal A}] &=& \int [d\mu(\xi)]_C \exp \left( 
ig \oint_C  \langle \Lambda |[V{\cal A} V^\dagger 
+ {i\over g} V d V^\dagger ]
|\Lambda \rangle \right)
\nonumber\\
&=& \int [d\mu(\xi)]_C \exp \left( 
ig \oint_C  [ n^A {\cal A}^A  
+ {1\over g}\omega ] \right) ,
\end{eqnarray}
where $[d\mu(\xi)]_C$ is the
product measure of
$d\mu(w(x),\bar w(x))$ along the loop, 
and 
\begin{eqnarray}
 n^A(x) := \langle \Lambda | V(x) T^A V^\dagger(x) |\Lambda \rangle .
\end{eqnarray}
Using the (usual) Stokes theorem
$
 \oint_{C=\partial S} \omega = \int_S d \omega  ,
$
we arrive at the NAST for $SU(3)$:
\begin{eqnarray}
 W^C[{\cal A}] = \int [d\mu(\xi)]_C \exp \left( 
ig \int_{S:\partial S=C} [ d(n^A {\cal A}^A) + {1\over g}\Omega_K ] \right) ,
\quad
\Omega_K := d\omega .
\label{NAST}
\end{eqnarray}
Taking into account
$
 d = \partial + \bar \partial  
 = dw_\alpha {\partial \over \partial w_\alpha} 
 + d\bar w_\beta {\partial \over \partial \bar w_\beta},
$
we find for $SU(3)$
\begin{eqnarray}
  \Omega_K &=&   d\omega 
  = im (\Delta_1)^{-2}[(1+|w_1|^2)dw_2 \wedge d\bar w_2
  - \bar w_2 w_1 dw_2 \wedge d\bar w_1
  \nonumber\\&&
  - w_2 \bar w_1 dw_1 \wedge d\bar w_2
  + (1+|w_2|^2) dw_1 \wedge d\bar w_1]
  \nonumber\\&&
  + in (\Delta_2)^{-2}[\Delta_1 dw_3 \wedge d\bar w_3
  - (w_1+\bar w_3 w_2) dw_3 \wedge (d\bar w_2 - \bar w_3 d\bar w_1)
    \nonumber\\&&
  - (\bar w_1+w_3 \bar w_2)(dw_2-w_3 dw_1) \wedge d\bar w_3
  \nonumber\\&&
  + (1+|w_3|^2)(dw_2-w_3 dw_1)(d\bar w_2 - \bar w_3 d\bar w_1)] .
  \label{K2f}
\end{eqnarray}
For $SU(2)$, we reproduce the well-known results;
$ \omega = i m {wd\bar w \over 1+|w|^2}   
$
and
$
 \Omega_K = i m (1+|w|^2)^{-2} dw \wedge d\bar w  .
$
\par
The flag manifold $F_{N-1}$ (including $CP^{N-1}$) is a K\"ahler manifold. 
So, it possesses complex {\it local} coordinates $w_{\alpha}$, an
Hermitian Riemannian metric
$ds^2=g_{\alpha\bar \beta}dw^{\alpha}d\bar w^{\beta}$, and a
corresponding two-form, 
$\Omega_K=i g_{\alpha\bar \beta}dw^{\alpha} \wedge d\bar w^{\beta}$
(K\"ahler form) which is closed, i.e., $d\Omega_K=0$. Any closed form
$\Omega_K$ is {\it locally} exact,
$\Omega_K= d\omega$ due to Poincar\'e's lemma, which is
consistent with (\ref{K2f}).
The metric $g_{\alpha\beta}$ is obtained as
$
 g_{\alpha\bar\beta} = {\partial \over \partial w^\alpha}
 {\partial \over \partial \bar w^\beta} K ,
$
from the K\"ahler potential $K=K(w,\bar w)$.  
Indeed, the $\Omega_K$ just obtained in (\ref{K2f}) agrees with the
K\"ahler two-form,
$
 \Omega_K = i \partial \bar \partial K 
$
obtained from the K\"ahler potential for $F_2$,
$
  K(w,\bar w) = \ln [(\Delta_1)^m(\Delta_2)^n] .
$
Hence, 
$
 \omega = {i \over 2}(\partial - \bar \partial) K , 
$
since
$
  \partial^2 = 0 = \bar \partial^2, 
  \partial \bar \partial + \bar \partial \partial = 0.
$
For CP$^2$, the K\"ahler potential 
$
  K(w,\bar w) = \ln [(\Delta_1)^m]
$
 is obtained as a special case of $F_2$ by putting $w_3=0$ and
$n=0$. For $F_1=CP^1$, 
$
  K(w,\bar w) = \ln [(1+|w|^2)^m], m=2j .
$

We define 
$
 n^A(x) := \langle \Lambda | V(x) T^A V^\dagger(x) |\Lambda \rangle
 = {1 \over 2}\bar \omega_a(x) (\lambda^A)_{ab} \omega_b(x),
$
where
$
  \omega_a(x) := V^\dagger(x) |\Lambda \rangle .
$
Especially, in the $CP^{2}$ case, we can write
\begin{eqnarray}
 n^A(x) := \langle \Lambda | U(x) T^A U^\dagger(x) |\Lambda \rangle
 = {1 \over 2}\phi_a^*(x) (\lambda^A)_{ab} \phi_b(x),
 \label{CPrep}
\end{eqnarray}
where $U\in SU(3)$ and 
$|\Lambda \rangle = (1,0,0)^t$. Then
$
 \phi_a := (U^\dagger(x) |\Lambda \rangle)_a
 = U_{1a}^*$ and hence $n^A = {1 \over 2}(U\lambda^A U^\dagger)_{11} .
$
On the other hand, we examine another expression (adjoint orbit
representation),
\begin{eqnarray}
  n^A={\rm tr}(U^\dagger {\cal H} U T^A) 
  = {1 \over 2}{\rm tr}({\cal H} U \lambda^A U^\dagger) ,
  \quad
 {\bf n} := n^A T^A = U^\dagger {\cal H} U ,
 \label{adjrep}
\end{eqnarray}
where we take
$
{\cal H} = \vec \Lambda \cdot (\lambda^3, \lambda^8)
= \Lambda_1 \lambda^3 + \Lambda_2 \lambda^8 .
$
For $[1,0]$ or $[0,1]$, we find
$
 {\cal H} = {\rm diag}({2 \over 3},-{1 \over 3},-{1 \over 3})  ,
$
or
$
 {\rm diag}({1 \over 3},-{2 \over 3},{1 \over 3})  ,
$
and two definitions (\ref{CPrep}) and (\ref{adjrep}) are equivalent,
$
n^A = {1 \over 2}{\rm tr}({\cal H} U \lambda^A U^\dagger)
={1 \over 2}(U\lambda^A U^\dagger)_{11}
$, 
since $U \lambda^A U^\dagger$ is traceless.
The $F_2$ variables $w_a$ and the
$CP^2$ variables $\phi_a$ are related as 
$\phi_1=1, \phi_2=w_1, \phi_2=w_2$ and
$\omega_a=\phi_a/\phi_1=w_{a-1}$ ($w_0:=1$ by definition).
In the $CP^{N-1}$ case, 
$
 \langle \Lambda |f(V) |\Lambda \rangle
 = 2{\rm tr}[{\cal H} f(U)] .
$
Hence,
$
 \omega(x) := 2{\rm tr}[{\cal H} iU(x) d U^\dagger(x)]
 = - i2{\rm tr}[{\cal H}dU(x)U^\dagger(x)]
$
which is a diagonal piece of the Maurer-Cartan one-form $dU U^{-1}$.
The two-form $\Omega_K$ is the symplectic two-form 
\cite{FN98},
$
 \Omega_K =  d\omega = 2{\rm tr}({\cal H}[U^{-1}dU,U^{-1}dU])
 = 2{\rm tr}({\bf n} [d{\bf n},d{\bf n}]).
$
\par
For $SU(3)$, the topological part $\int_S \Omega_K=\oint_C \omega$
corresponding to the residual $U(2)$ invariance is interpreted as the
geometric phase of the  Wilczek-Zee holonomy \cite{WZ84}, just as it
is interpreted in the $SU(2)$ case as the Berry-Aharonov-Anandan phase
for the residual $U(1)$ invariance. The details will be given in a
subsequent paper
\cite{KT99}.

\section{Abelian dominance and area law}
\par
The NAST (\ref{NAST}) implies that the expectation value of the
Wilson loop in the SU(N) Yang-Mills theory is given by
\begin{eqnarray}
 \langle W^C[{\cal A}] \rangle_{YM}
 =  \Big\langle \exp \left( 
ig \int_{S:\partial S=C} [ da] \right)
\Big\rangle_{YM}  
 =  \Big\langle \exp \left( 
ig \int_{S} [d(n^A {\cal A}^A) + {1\over g}\Omega_K ] \right)
\Big\rangle_{YM} ,
\end{eqnarray}
where the one-form $a$ is written as
\begin{eqnarray}
 a := n^A {\cal A}^A  + {1\over g}\omega 
 = \langle \Lambda | {\cal A}^V |\Lambda \rangle .
 \label{ap}
\end{eqnarray}
Here ${\cal A}^V:=V{\cal A} V^\dagger + {i \over g}V d V^\dagger $ 
is the gauge transformation of ${\cal A}$ by
$V \in F_{N-1}$.  For quark in the fundamental representation,
$
 a  = 2{\rm tr}({\cal H}{\cal A}^V) .
$
So, the one-form $a$ is equal to the diagonal
piece of the gauge-transformed potential ${\cal A}^V$.
In the $SU(2)$ case, $a={\rm tr}(T^3{\cal A}^V)$ for any
representation and the two-form $f:=da$ is the Abelian field strength
which is invariant under the $SU(2)$ transformation. The two-form $f$
is nothing but the 't Hooft tensor 
$f_{\mu\nu}$ describing the magnetic flux
emanating from the magnetic monopole,
$
 f_{\mu\nu}(x) :=  \partial_\mu(n^A(x){\cal A}_\nu^A(x)) 
   - \partial_\nu(n^A(x){\cal A}_\mu^A(x))
   - {1 \over g}{\bf n}(x) \cdot (\partial_\mu {\bf n}(x) \times
\partial_\nu {\bf n}(x)) ,
$
if we identify $n^A$ with the
direction of the Higgs field,
$
 \hat \phi^A :=\phi^A/|\phi|, \quad |\phi|:= \sqrt{\phi^A \phi^A} . 
$
In general, the (curvature)
two-form 
$f=d(n^A {\cal A}^A) + {1\over g}\Omega_K$ in the NAST is the Abelian field
strength which is invariant even under the non-Abelian gauge
transformation of
$G$.  It reduces to the 't Hooft tensor in $SU(2)$ case. 

\par
The Abelian dominance in $SU(N)$ Yang-Mills theory is derived as
follows.   We adopt the maximal Abelian (MA) gauge.
The MA gauge for $SU(N)$ is defined as follows.
Under the Cartan decomposition of ${\cal A}$ into the diagonal
($H$) and off-diagonal ($G/H$) pieces,
$
 {\cal A} = a^i H^i + A^a T^a ,
$
the MA gauge condition is given by
$
 \partial_\mu A_\mu^a - g f^{abi}a_\mu^i A_\mu^b = 0 ,
$
which is obtained by minimizing the functional
$
 {\cal R} := \int d^4x {1 \over 2} A_\mu^a A_\mu^a 
 := \int d^4x {1 \over 2} {\rm tr}_{G/H}({\cal A}_\mu{\cal A}_\mu),
$
under the gauge transformation.
For $SU(3)$,
$
 H^1={\lambda^3 \over 2}, H^2={\lambda^8 \over 2}, 
 T^a = {\lambda^a \over 2} \ (a=1,2,4,5,6,7).
$
The low-energy effective gauge theory of QCD has been
derived in the MA gauge
by integrating out the
off-diagonal gauge fields (together with the ghost and anti-ghost
fields)
\cite{QR97,KondoI}.  
Then the $SU(N)$ Yang-Mills theory has been reduced to the Abelian
gauge theory with the gauge coupling  $g$ which runs according to
the same renormalization-group beta function as the original $SU(N)$
Yang-Mills theory,
\begin{eqnarray}
  \Big\langle \cdots \Big\rangle_{YM} 
  &=& Z_{YM}^{-1} \int {\cal D}a \exp (iS_{APEGT}[a]) \cdots ,
  \\
  S_{APEGT}[a] &=& -\int d^4 x {1 \over 4g^2(\mu)} (da, da) ,
  \label{action1}
  \\
  {1 \over g(\mu)^2} &=&  {1 \over g(\mu_0)^2}
  + {b_0 \over 8\pi^2} \ln {\mu \over \mu_0}, \quad
  b_0 = {11n \over 3} > 0 ,
  \label{runc}
\end{eqnarray}
up to the
one-loop level.
\footnote{
The result of \cite{QR97,KondoI} for $SU(2)$ can be generalized to
$SU(N)$ in the straightforward way, at least in one-loop level
\cite{KS99}. In the two-loop level, it is not trivial.  The two-loop
result will be given elsewhere \cite{KS99}.
}
This result \cite{KondoI} combined with the $SU(2)$ NAST
\cite{DP89,KondoIV} implies the Abelian dominance in the low-energy
region of $SU(2)$ QCD, as shown in
\cite{KondoIV}. By virtue of the NAST for $SU(3)$ just
derived, Abelian dominance in $SU(3)$ Yang-Mills theory follows
immediately from the same argument as above,
if we identify the connection one-form $a$ with the diagonal
piece of the gauge potential ${\cal A}^V$ given by (\ref{ap}). 
The monopole dominance is more subtle.
It was derived for $SU(2)$ in
\cite{KondoIV} by showing that the dominant contribution to the
area law comes from the monopole piece alone,
$
 \Omega_K  = d\omega = 2{\rm tr}({\bf n} [d{\bf n},d{\bf n}]).
$
\par
In the following derivation of quark confinement, the magnetic
monopoles (equivalently, the instantons in the coset $F_{N-1}$ NLSM)
is expected to give the dominant configuration to the area law of the Wilson
loop or the string tension.      
The maximal stability group $\tilde{H}$ should be distinguished from the residual gauge group
$H$ retained after the partial gauge fixing $G \rightarrow H$ which realizes the
magnetic monopole. 
The existence of magnetic monopole is suggested
from the non-trivial Homotopy groups $\pi_2(G/\tilde{H})$.
In the case (II),
$
 \pi_2(SU(3)/(U(1) \times U(1)))=\pi_1(U(1) \times U(1)) ={\bf
Z}+{\bf Z} .
$
On the other hand, in the case (I) 
$
 \pi_2(SU(3)/U(2))=\pi_1(U(2))=\pi_1(SU(2)\times
U(1))=\pi_1(U(1))={\bf Z} .
$
Note that $CP^{N-1}$ NLSM has only the local $U(1)$ invariance for any
$N\ge 2$.  It is this $U(1)$ invariance that corresponds to a kind of
Abelian magnetic monopole appearing in the case (I).
This situation should be compared with the $SU(2)$ case where the
maximal stability group $\tilde{H}$ is always given by the maximal torus $H=U(1)$
irrespective of the representation.  Therefore, the coset is given by
$G/\tilde{H}=SU(2)/U(1)=F_1=CP^1 \cong  S^2  $ for any
representation and $\pi_2(SU(2)/U(1))={\bf Z}$.
For $SU(3)$, our results suggest that  the fundamental quarks are  to
be confined even when the residual gauge group $H$ is given by
$\tilde{H}=U(2)$ and $\pi_2(G/H)={\bf Z}$, while the adjoint quark is
related to the maximal torus
$H=U(1)\times U(1)$ and
$\pi_2(G/H)={\bf Z}+{\bf Z}$.
This observation is in sharp contrast with the
conventional treatment in which the $(N-1)$ species of magnetic
monopoles corresponding to the residual maximal torus group
$U(1)^{N-1}$ of $G=SU(N)$ are taken into account on equal footing. 
\par
We adopt an special choice of the MA gauge for fixing the gauge also
in the $SU(3)$ case in which the gauge fixing part is given by
$
 S_{GF} = \int d^4 x \ i\delta_B \bar \delta_B {\rm tr}_{G/H}
\left[ 
 {1 \over 2}\Omega_\mu \Omega_\mu + i C \bar C 
 \right] 
$
where
$
 \Omega_\mu := {i \over g} U \partial_\mu U^\dagger ,
$ 
with the BRST $\delta_B$ and anti-BRST $\bar \delta_B$
transformations, see \cite{KondoII} for details.
We reformulate the Yang-Mills theory as a
perturbative deformation of the topological quantum field theory
(TQFT) which is obtained from the gauge fixing part for the compact
gauge variable \cite{KondoII,KondoVI}.  Repeating the similar
arguments given in
\cite{KondoII,KondoIV,KondoVI}, we can show
\cite{KT99} that  the area decay of the Wilson loop 
$\langle W^C[{\cal A}] \rangle_{YM}$
in the four-dimensional Yang-Mills
theory is dominated by the diagonal Wilson loop 
$
 \langle \exp \left( i \int_{S} \Omega_K \right) \rangle_{TQFT_4} 
$ 
in the four-dimensional TQFT with an action $S_{GF}$ which describes
the magnetic monopole in four dimensions. When the Wilson loop is
planar, it is shown that this expectation value is equal to the the
instanton distribution
$
 \langle \exp \left( i \int_{S} \Omega_K \right) \rangle_{NLSM_2} 
$ 
in the two-dimensional coset G/H NLSM with an action $S_{NLSM}$. This is due
to dimensional reduction. The dimensional reduction leads to the two-dimensional
coset
$F_{N-1}=SU(N)/U(1)^{N-1}$ NLSM,
\begin{eqnarray}
  S_{NLSM} = {4\pi \over g^2(\mu)} \int d^2x g_{\alpha\bar \beta}
  {\partial w^{\alpha} \over
\partial x_a}{\partial \bar w^{\beta} \over \partial x_a} 
= {4\pi \over g^2(\mu)} \int dzd\bar z \ g_{\alpha\bar \beta} 
 \left(
 {\partial w^\alpha \over \partial z}{\partial \bar w^\beta \over
\partial \bar z}
  + {\partial w^\alpha \over \partial \bar z}{\partial \bar w^\beta
\over \partial z}
  \right) ,
\end{eqnarray}
where $g(\mu)$ is the running Yang-Mills coupling constant
(\ref{runc}) induced from the perturbative deformation piece in four
dimensions. Thus the derivation of the area law of the Wilson loop in
the four-dimensional Yang-Mills theory is reduced to that of the
diagonal Wilson loop in the two-dimensional coset NLSM.

\par
The area law of the Wilson loop is shown
as follows. The finite action configuration of the coset NLSM is
provided with the instanton solution.  
For instanton configuration, 
$
 S_{NLSM} = {4\pi^2 \over g^2}|Q| ,
$
with a topological charge $Q$.
It is known \cite{Perelomov87}
that the topological charge $Q$ of the instanton in the $F_{N-1}$ NLSM 
$(N\ge 3)$ is
given by the integral of the K\"ahler 2-form,
\begin{eqnarray}
 Q = {1 \over \pi} \int \Omega_K 
= \int_{{\bf R}^2} {d^2x \over \pi} 
\epsilon_{ab} g_{\alpha\bar \beta}
  {\partial w^{\alpha} \over
\partial x_a}{\partial \bar w^{\beta} \over \partial x_b} 
=  \int_{{\bf C}} {dzd\bar z \over \pi} \ g_{\alpha\bar \beta} 
 \left(
 {\partial w^\alpha \over \partial z}{\partial \bar w^\beta \over
\partial \bar z}
  - {\partial w^\alpha \over \partial \bar z}{\partial \bar w^\beta
\over \partial z}
  \right) .
\end{eqnarray}
This is a generalization of the $F_1$ case,
\begin{eqnarray}
  Q = {i \over 2\pi}\int_{{\bf C}} {dwd\bar w \over (1+|w|^2)^2}
  =  {i \over 2\pi}\int_{S^2} {dzd\bar z \over (1+|w|^2)^2}
 \left(
 {\partial w \over \partial z}{\partial \bar w \over \partial \bar z}
  - {\partial w \over \partial \bar z}{\partial \bar w \over \partial
z}
  \right) .
\end{eqnarray}
Since the K\"ahler potential for $F_n$ is written as
\begin{eqnarray}
 K(w,\bar w) = \sum_{\ell=1}^{n} d_\ell K_\ell(w,\bar w)
 = \sum_{\ell=1}^{n} d_\ell \ln \Delta_\ell (w,\bar w) 
\end{eqnarray}
with the Dynkin indices $d_\ell (\ell=1,\cdots,n)$, the integral
$\int \Omega_K$ over the whole two-dimensional space reads
$
 \int_{{\bf R}^2} \Omega_K = \pi Q 
 = \pi \sum_{\ell=1}^{n} d_\ell
Q_\ell
$
where $Q_\ell$ are integers-valued topological charges.
Hence, $\Omega_K(x)/\pi$ is identified with the density of the
topological charge (up to the weight due to the index $d_\ell$).
Then, for the large Wilson loop compared with the typical size of
the instanton,
$\int_S \Omega_K(x)/\pi$ in the NAST counts the number of instantons
minus anti-instantons which are contained inside the area 
$S \subset {\bf R}^2$
bounded by the loop $C$.  Thus, the expectation value 
$
 \langle \exp \left( i \int_{S} \Omega_K \right) \rangle_{NLSM} 
 = Z_{NLSM}^{-1} \int [d\mu(w,\bar w)] \exp (-S_{NLSM}[w,\bar w]) 
 \exp \left( i \int_{S} \Omega_K \right)
$ 
is calculated by summing over all the possibilities of instanton and
anti-instanton configurations or integration over the instanton
moduli.  For the quark
in the fundamental representation ($d_1=1,d_2=d_3=\cdots =d_n=0$),
this is easily performed as follows. 
Especially, in the $SU(3)$ case with $[1,0]$, $\xi$ is independent
of $w_3$, so, $w_3$ is redundant in the $F_2$ NLSM.  Hence, it
suffices to consider the $CP^2$ NLSM for the fundamental quark. If we put
$w_2=0$, then
$  \Omega_K  
  = i  (1+|w_1|^2)^{-2} dw_1 \wedge d\bar w_1 .
$
Similarly, if $w_1=0$, then
$  \Omega_K  
  = i  (1+|w_2|^2)^{-2} dw_2 \wedge d\bar w_2  .
$
For a polynomial $w_\alpha=w_\alpha(z)$  in $z=x+iy$ with an order
$n$, we find an instanton charge,
$\int \Omega_K = \pi Q$, $Q=n$.
This is the same situation as that encountered in
$SU(2)$, $\int \Omega_K =2j \pi Q$ $(j=1/2)$
\cite{KondoII,KondoIV}.
Thus the Wilson loop is estimated by the instanton calculus.  
In fact, the dilute instanton gas approximation leads to
the area law for the Wilson loop, see \cite{KondoII}.  
This calculation is improved by including fluctuations from the
instanton solutions and this issue will be discussed in
detail elsewhere. 
In the paper \cite{KT99} we have applied the large N expansion to 
obtain the systematic result. At least in the leading order in 1/N expansion, 
we have obtained the result in consistent with instanton calculas 
 for Wilson loop.
\par
Finally, some comments are in order.
\par
1)
We want to emphasize that we have actually
"derived" the area law of the Wilson loop by calculating the
contributions from magnetic monopole.  This is based on the
combination of the non-Abelian Stokes theorem (NAST) and the
reformulation of Yang-Mills theory within the framework of
perturbative deformation. In other words, we have proposed a picture
in which our framework is self-consistent.

Of course, we have obtained these results within the framework in
which the quantum fluctuations around the magnetic monopole are
treated in the perturbative way (which we called the perturbative
deformation).  However, this is the first analytical result of having
derived the area law from the first principle by taking into account
magnetic monopole in "3+1" dimensions.  
Our results should be regarded as the monopole dominance in the
weak sense.  That is to say, the magnetic monopole contribution is
enough to derive the area law.  The monopole dominance can be
a sufficient condition for quark confinement, even if it is not a
necessary condition

\par
2)
It should be remarked that the maximal
stability group $\tilde H$ is in general different from the residual
gauge group
$H$ for partial gauge fixing $G \rightarrow H$ and that
$\tilde H$ is larger than $H$, i.e., $H
\subset
\tilde H$. For the coset, therefore, we have $G/\tilde H \subset
G/H$.  For
$G=SU(3)$,
$\tilde H=U(2)$ or $U(1) \times U(1)$ depending on the
representation in question.
By making use of the NAST, the Wilson loop is rewritten in terms of
the $G/\tilde H$ variables.  On the other hand, after partial gauge
fixing $G \rightarrow H$,  the gauge-fixing part is rewritten into
the NLSM in terms of the $G/H$ variables.  So, the NLSM involves
redundant variables, since $G/\tilde H \subset G/H$.
Thus our claim is that the degrees of freedom which are directly
responsible to the area law are given by $G/\tilde H$ rather than
$G/H$. For the fundamental representation, 
$G/\tilde H=SU(3)/U(2)=CP^2$ and $CP^{N-1}$ model has local gauge
invariance $U(1)$ for any $N \ge 2$.  In this sense, we can adopt
even $H=U(2)$ in the partial gauge fixing at least for the
fundamental quark.  In this case, $G/\tilde H = G/H$.
Thus, the resultant theory after partial gauge fixing is not
necessarily a U(2) gauge theory and we don't intend to claim the
equality of the string tensions between U(2) and SU(3).
Rather, the two string tensions for $H=U(2)$ and $H=U(1)\times U(1)$
should coincide with each other.

\par
3)
Baryons are also important physical objects
from the viewpoint of color confinement in the SU(3) gauge theory
as well as mesons.
In the MA gauge, the remaining gauge symmetry is
the local $U(1)^2$ symmetry and the global Weyl symmetry.
The Weyl symmetry is a remanant of SU(3), owing to which one will
find only the color-singlet combination of quarks.
In the presentation of our paper, however, there is no Weyl
symmetry in a "manifest" form, because we can select the reference
state from any of three sates and we have selected a special highest
weight vector denoted by
$\Lambda$ for the fundamental representation.  
Once a state is selected, one complex coordinate in $F_2$ becomes
irrelevant  in the expectation value of the Wilson loop and
the relevant space becomes $CP^2$ which is spanned by other
two complex coordinates in $F_2$.  The change of reference
state causes only the change of irrelevant coordinate and the
Wilson loop has the same expectation value. Therefore, the Weyl
symmetry is easily recovered by averaging over all possible choices
of the highest weight vector.  The details will be given elsewhere.
\section*{Acknowledgments}
This work is supported in part by
the Grant-in-Aid for Scientific Research from the Ministry of
Education, Science and Culture (No.10640249).


\end{document}